\documentstyle[aaspptwo]{article}
\newcommand{\Xray}{\mbox{X-ray}}
\begin{document}
\title{Optical and Infrared Observations of SGR~1806$-$20\footnote{%
This is a preprint of a paper submitted to {\em The Astrophysical Journal
(Letters)}. No bibliographic reference should be made to this preprint.
Permission to cite material in this paper must be received from the
authors.}}
\author{S. R. Kulkarni, K. Matthews,  G. Neugebauer,\\
        I. N. Reid, M. H. van Kerkwijk, G. Vasisht}
\affil{Palomar Observatory, California Institute of Technology
105-24, Pasadena, CA 91125}
\begin{abstract}
The soft gamma-ray repeater (SGR) 1806$-$20 is associated with the
center-brightened non-thermal nebula G~10.0$-$0.3, thought to be a
plerion.  As in other plerions, a steady \Xray\ source,
AX~1805.7$-$2025, has been detected coincident with the peak of the
nebular radio emission.  Vasisht et al.\ have shown that the radio
peak has a core-jet appearance, and argue that the core marks the true
position of the SGR.  At optical wavelengths, we detect three objects
in the vicinity of the radio core.  Only for the star closest to the
core, barely visible in the optical but bright in the infrared
($K=8.4\,$mag.), the reddening is consistent with the high extinction
($A_V\simeq30\,$mag.) that has been inferred for AX~1805.7$-$2025.
{}From the absence of CO band absorption, we infer that the spectral
type of this star is earlier than late~G/early~K.  The large
extinction probably arises in a molecular cloud located at a distance
of 6$\,$kpc, which means that the star, just like AX~1805.7$-$2025, is
in or behind this cloud.  This implies that the star is a supergiant.
Since supergiants are rare, a chance coincidence with the compact
radio core is very unlikely.  To our knowledge, there are only three
other examples of luminous stars embedded in non-thermal radio
nebulae, SS~433, \mbox{Cir X-1} and G~70.7+1.2.  Given this and the
low coincidence probability, we suggest that the bright star is
physically associated with SGR~1806$-$20, making it the first stellar
identification of a high-energy transient.
\end{abstract}

\keywords{gamma rays: bursts --
          stars: individual: SGR~1806$-$20 --
          stars: neutron --
          X rays: stars}

\section{Introduction}
The three known soft gamma-ray repeaters constitute a sub-class of
high-energy transients marked by distinctive temporal and spectral
features (see Higdon \& Lingenfelter \cite{higdl:90} for a review).
SGR 1806$-$20 is of particular interest because it is located in our
Galaxy and has been identified with a radio supernova remnant,
G~10.0$-$0.3 (Kulkarni \& Frail \cite{kulkf:93}, Murakami et al.\
\cite{mura&a:94}).  Based on this association, and  that of
SGR~0526$-$66 with N~49 (Cline et al.\ \cite{clin&a:82}; Rothschild,
Kulkarni, \& Lingenfelter \cite{roth&a:94}), it has been suggested
that young neutron stars produce the soft $\gamma$-ray outbursts
(Murakami et al.\ \cite{mura&a:94}; Kulkarni et al.\
\cite{kulk&a:94}).

However, it remains mysterious how a mere isolated neutron star can
produce the observed impressive burst luminosities, ranging from
$10^{41}$ to $10^{45}\,$erg$\,$s$^{-1}$.  For this reason, binary
models, specifically models appealing to accretion from a companion
(including comets; see Liang \& Petrosian \cite{lianp:86}), have
considerable appeal.  The simplest expectation of accretion models is
the presence of a mass-transferring secondary.  Independently, purely
on phenomenological grounds, Kulkarni et al.\ (\cite{kulk&a:94}) have
drawn attention to the similarity between SGR~1806$-$20/G~10.0$-$0.3
and the accreting neutron-star binary \mbox{Cir~X-1}, which is also
embedded in a non-thermal radio nebula.  Clearly, optical and infrared
(IR) observations can shed more light on the stellar configuration of
SGRs.

SGR~1806$-$20 is best suited to start looking for optical/IR
counterparts.  This is because there is not only excellent positional
coincidence between the $\gamma$-ray (Hurley et al.\
\cite{hurl&a:94}), \Xray\ (Murakami et al.\ \cite{mura&a:94}, Cooke
\cite{cook:93}) and radio positions (Kulkarni et al.\
\cite{kulk&a:94}, Vasisht et al.\ \cite{vasi&a:94}; see Table~1), but
also the association of the \Xray\ source with the SGR has been
strengthened by the observations of an \Xray\ burst by ASCA (Murakami
et al.\ \cite{mura&a:94}) that occurred simultaneously with a
$\gamma$-ray burst detected by BATSE (Kouveliotou et al.\
\cite{kouv&a:94}).  Murakami et al.\ found that at the position of the
\Xray\ burst, there was also a quiescent \Xray\ source,
AX~1805.7$-$2025.

The best localization of AX~1805.7$-$2025 is that by ROSAT (Cooke
\cite{cook:93}), with an error radius of 11 arcsec;
here and elsewhere the error radius is the 90\% probability radius.
This is still too coarse to make meaningful optical or IR
identifications, especially since the source is located in the
Galactic plane.  However, for SGR~1806$-$66 we can obtain a more
accurate position from radio observations.  This is because the
remnant G~10.0$-$0.3 does not have the usual shell morphology, but is
center-brightened, like a plerion (Kulkarni et al.\ \cite{kulk&a:94}).
Plerions are thought to be synchrotron bubbles powered by the
relativistic wind of a central compact object, usually a pulsar (see
Weiler \& Sramek \cite{weils:88} for a review).  The compact object is
almost always associated with an \Xray\ source.  Since the lifetime of
X-ray producing relativistic particles is short ($\sim$ years), the
\Xray\ sources are expected to mark the true location of the compact
source.  In most plerions, the radio emission also peaks on the \Xray\
point source.  Given the similarity of G~10.0$-$0.3/AX~1805.7$-$2025
to other plerions, we can therefore safely assume that
AX~1805.7$-$2025 marks the true location of SGR~1806$-$20, and that
radio observations can be used to further refine this position which
then enable us to search for optical/IR counterparts.

{}From the ASCA observations of AX~1805.7$-$2025, an absorbing hydrogen
column density of $(6\pm0.2)\times10^{22}\,{\rm{}cm}^{-2}$ is inferred
(Sonobe et al.\ \cite{sono&a:94}).  If this is due to interstellar
extinction then using the usual relation between $N_{\rm{}H}$ and
$E_{B-V}$ (Spitzer \cite{spit:78}), we derive $A_V=30\,$mag.  There is a
giant molecular cloud located at about 6$\,$kpc distance in this
direction (T.~Dame as quoted in Grindlay \cite{grin:94}), with an
inferred hydrogen column density quite similar to that discussed
above.  We note that since there are only two SGRs in the Galaxy
(Kouveleiotou et al.\ \cite{kouv&a:94}), the expected distance to SGR
1806$-$20 is $\sim\!R_0\simeq10\,$kpc.  Based on this very general
argument, as well as on the observation that the intrinsic column
densities in Galactic \Xray\ sources seldomly are so high, we would
argue that the source is in or behind the cloud and that most of the
column density is due to it.  Thus, we expect any plausible
counterpart to be heavily reddened.

Here, we first briefly discuss the best radio position currently
available, and then report on the results of a series of optical/IR
observations carried out at the Palomar Observatory.

\section{The Precise Position of SGR~1806$-$20}

In a companion paper, Vasisht et al.\ (1994) present a new high
angular resolution image of the core of G~10.0$-$0.3 made from data
obtained at the Very Large Array (VLA).  They show that the peak
region consists of a compact source with asymmetric linear emission to
the east.  Based on the similarity of the geometry with that seen in
extragalactic radio sources and Galactic accreting binaries, they
suggest that the asymmetric emission represents the flow of a
relativistic wind emanating from the compact source.  If the above
reasoning is correct, then the core represents the location of the
compact object.  We refer to this position of the core as the VLA-HR
(for high resolution) position.  Note that the VLA-HR position is
different from that derived earlier from low-resolution (beam of 6
arcsecond) VLA images (VLA-LR, Table~1; Kulkarni et al.\
\cite{kulk&a:93a}, \cite{kulk&a:94}).  The VLA-LR position was likely
biased by the east-west asymmetry in the structure of the source
(Vasisht et al.\ \cite{vasi&a:94}).

\section{Observations at Palomar}

I-band CCD images were obtained with the Palomar 60-inch telescope on
1993 October 4--5. On 1993 October 8, IR observations were conducted.
Images in the near-infrared J~(1.25\,\micron), H~(1.65\,\micron),
K~(2.2\,\micron), and L$'$~(3.7\,\micron) bands were obtained with a
Santa Barbara Research Corporation $58\times 62$ InSb array at the
$f/70$ Cassegrain focus of the Hale 200-inch telescope (Figure~1).
Grism spectra of `star~A' (see Figure~1) were taken with a slit width
of $1''$ and a resulting spectral resolution of $\sim\!75$. The
integration time for object and sky spectra was 100$\,$s. Atmospheric
and instrumental features were removed by dividing the spectrum of the
object by that of a G~dwarf (HR~7643) taken at the same airmass.

On 1994 June 1, we used the 4-SHOOTER, a CCD imager/spectrograph at
the Cassegrain focus of the Hale 200-inch telescope (Gunn et al.\
\cite{gunn&a:87}), to obtain Gunn $r$, $i$ and $z$ images of the same
field (see Figure~1).  In the same run, we also obtained a 45-minute
red spectrum of the nearby, relatively bright `star~I' (see Figure~1),
as well as, simultaneous with the spectrum, a 35-minute narrow-band
($20$\,\AA) H$\alpha$ image of the field surrounding the slit.

The positions of 16 stars visible on both the 60-inch I-band CCD image
and the appropriate Palomar Observatory Sky Survey plate were
determined relative to nearly 50 SAO stars using a measuring engine at
the Observatories of the Carnegie Institution of Washington.  The rms
residual was 0\farcs8 in both right ascension and declination.  With
the positions of the 16 stars thus determined, the position of star~I
was derived.  The position of star~I was also derived by relating the
9 out of 16 stars that were visible on a short Gunn-$i$ 4-SHOOTER
image.  These two methods yielded the same result to within 0\farcs2.
Separately, we determined the position of star~I with respect to two
stars in the Guide Star Catalog (GSC; Russell et al.\
\cite{russ&a:90}) that were visible on the 4-SHOOTER images, using the
known orientation and pixel scale.  This determination was also in
excellent agreement with the others (Table~1).  For the astrometry of
the Gunn $i$ and $z$ images, we determined the positions of 15 stars
visible on the short exposure and not (too) overexposed on the longer
ones.  Subsequently, the J-band astrometry was done relative to the
Gunn-$z$ image using stars 1--5 and~I, and that of the H and K~bands
relative to~J with stars 1, 4, 6, and~A (see Figure~1, Table~1).

In the $z$-band image, two objects are visible close to the radio
position(s), `I', a star, and `HK', a fuzzy object (see Figure~1).
The shape of HK is consistent with it being a pair of stars, aligned
roughly east west.  We will refer to the western portion as~H and the
eastern portion as~K. Both parts are barely visible in the $r$ frame
suggesting that they are highly reddened.  Star~I is moderately
reddened.

In the J and K-band images, two IR stars are seen close to the radio
position, star~A and the infrared counterpart of star~I mentioned
above.  From the astrometry (Table~1), it is clear that star~A is
most likely coincident with the western part~H of the HK fuzz
(consistent with the fact that this part gets brighter going from $i$
to $z$ band).  Based on crude astrometry with the telescope, we had
earlier concluded in an IAU telegram (Kulkarni et al.\
\cite{kulk&a:93b}) that star~A was the IR counterpart of star~I. From
Figure 1, it is clear that stars A and I are manifestly different
stars.

For star~A, we obtained the following magnitudes: 13.3~[J], 10.2~[H],
8.4~[K] and 7.0~[L$'$].  (Note that this K-band value supersedes that
given in Kulkarni et al.\ (\cite{kulk&a:93a}).  That value was
affected by saturation in the K-band data.)  From our optical data, we
find for the HK fuzz magnitudes of $>$24.4~[$r$], 21.6~[$i$] and
20.5~[$z$].  In the $z$ band, the contribution of star~A (part~H of
the fuzz) is about 60\% ($z\simeq20.8\,$mag.), while in the $i$ band
it is $\la30$\% ($i\ga22.5\,$mag.).  We note that the Gunn magnitudes
of HK should be considered rough estimates, since they may be quite
severely influenced by color terms resulting from the interstellar
extinction (e.g., Grebel \& Roberts \cite{grebr:94}).  For star~I we
find magnitudes of 21.5~[$r$], 20.0~[$i$], and 19.4~[$z$] in the
optical, and 16.3 in~J.

{}From the 2100-s exposure taken with the H$\alpha$ filter, we derive,
after a scaled subtraction of an $r$-band image, a 1-$\sigma$ upper
limit of 50 photoelectrons per $0\farcs33\times0\farcs33$ pixel in the
vicinity of the HK fuzz. We have estimated the total throughput
(telescope, camera, H$\alpha$ filter) to be 10\%. We thus derive a
1-$\sigma$ upper limit to an observed H$\alpha$ surface brightness of
$0.46\times10^6\,$photons$\,$cm$^{-2}\,$s$^{-1}\,$sterad$^{-1}$ or 6\,R
(Rayleigh). The emission measure 1-$\sigma$ limit on an angular scale
$\theta$, is $16(\theta/0\farcs33)^{-1}\,{\rm cm^{-6}\,pc}$.  While this
is a very good limit, any H~II region associated with SGR 1806$-$20
would suffer from severe extinction (see Section~1) and consequently
the intrinsic limit on the presence of an H~II region is considerably
worse.

\section{Discussion\label{sec:disc}}
Given the complexity of the radio emission, it is important to
consider all sources in the vicinity of the VLA-HR position as
potential SGR counterparts. This means stars~I, A and part K of the HK
fuzz (part~H is the optical counterpart of star~A; see Section 3).
The optical spectrum of star~I shows that it is a moderately reddened
object with no obvious emission lines. We consider this to be an
unlikely counterpart because ({\em{}i}) it is outside the nominal 90\%
astrometric error radius (Table 1), ({\em{}ii}) the reddening seen in
the optical spectrum is not as large as that inferred for
AX~1805.7$-$2025 and ({\em{}iii}) it has no unusual features in the
spectrum (i.e., bright emission lines).  However, as pointed out by
the referee, there is a considerable range in the emission line
strengths of accreting binaries.  A more sensitive spectrum at higher
spectral resolution is needed to firmly rule out that star I has
emission lines at all.  At this time we have little information about
star~K, but we do note that it too, like star I, is not bright in the
near IR bands suggesting that it is not very highly reddened.

{}From positional and extinction considerations, star~A is the most
likely candidate.  Using the Galactic star count model of Jones et
al.\ (\cite{jone&a:81}), we estimate that the probability of a chance
coincidence of an IR star as bright as or brighter than star~A
(K$\le8.4\,$mag.) with a patch, 3-arcsec on a side, is only
$10^{-3}$. This is small enough to warrant further discussion.

The K-band grism spectrum shows no features that we can readily
identify.  Most notable is the absence of the band structure of CO
beyond 2.3$\,$\micron, a feature which is common in the spectra of
luminous late-type stars (Kleinmann \& Hall \cite{kh:86}; Lancon \&
Rocca-Volmerange \cite{lancr:92}).  Thus, the star must be earlier
than spectral type late~G/early~K.

With two colors ($J-H$, $H-K$) we can constrain the extinction to the
star, as well as, to some extent, its spectral type.  Assuming an
interstellar extinction law appropriate to the Galactic Center (Rieke
\&\ Lebofsky \cite{riekl:85}) we find (1) $A_V\simeq30\,$mag.\ and (2)
dereddened IR magnitudes of 5.8~[L$'$], 5.39~[K], 5.41~[H], and
5.79~[J]. The colors $J-H=0.38\,$mag.\ and $H-K=0.02\,$mag.\ are
appropriate for a late G star.  We recognize that inference (2) is
prone to small deviations in the assumed shape of the reddening law.
In particular, we cannot rule out that the star is earlier than
spectral type~G.  However, since the intrinsic IR colors of a star
earlier than spectral type G are essentially neutral
($J-H\simeq0\,$mag., $H-K\simeq0\,$mag.), inference (1) is on a firm
basis.  Therefore, star~A, like AX~1805.7$-$2025, is most likely in or
behind the giant molecular cloud that is seen at about 6\,kpc in this
direction (Section~1).  Thus, the absolute K magnitude of star~A is
$<\!-8.5$, i.e., star~A is a supergiant.  According to the star-count
model of Jones et al.\ (\cite{jone&a:81}), most of the luminous stars
in this direction are cool (K-M) giants.  The chance coincidence
probability of a hot supergiant with the radio position is exceedingly
small, $\sim\!2\times10^{-5}$.

We conclude from the positional coincidence, the consistency of
reddening estimates, and the low a posteriori probability of a chance
coincidence that star~A is almost certainly the infrared counterpart
of the core of the radio emission, and thus, a very good candidate for
being the stellar counterpart of SGR~1806$-$20.  We end by noting
that, to our knowledge, there are only three other luminous stars
embedded in non-thermal radio nebulae, SS~433 (Margon \cite{marg:84}),
\mbox{Cir~X-1} (Stewart et al.\ \cite{stew&a:93}) and G~70.7+1.2
(Kulkarni et al.\ \cite{kulk&a:92}), all of which also show quiescent
\Xray\ emission.  The similarity of
star~A/AX~1805.7$-$2025/G~10.0$-$0.3 to these systems is quite
striking.  Purely on phenomenological grounds, therefore, further
investigations of star~A are warranted.

\acknowledgements
We thank D.~Frail for advance information of the radio data and
T.~Nakajima for help with evaluating the chance coincidence
probabilities.  We gratefully acknowledge discussions with D.~Frail
and T.~Murakami, and the referee, B.~Margon, for suggestions which
resulted in a better flow of logic.  Some of the optical observations
were made at the Palomar 60-inch telescope which is jointly owned by
the California Institution of Technology and the Carnegie Institute of
Washington.  SRK's research is supported by the US NSF, NASA and the
Packard Foundation. MHvK is supported by a NASA Hubble Fellowship.
Infrared astrophysics at Caltech is supported by grant from the NSF.

\clearpage
\thispagestyle{empty}
\newcommand{\pha}{\mbox{$\phantom{\mbox{18 05 }}$}}
\newcommand{\phd}{\mbox{$\phantom{\mbox{$-$20 25 }}$}}
\def\tablevspace#1{\noalign{\vskip #1}}
\begin{planotable}{lllll}
\tablewidth{0pc}
\tablecaption{Positions of Counterparts and Nearby Stars}
\tablehead{
\colhead{Name}&
\colhead{$\alpha(1950)$}&
\colhead{$\delta(1950)$}&
\colhead{$r_{90}$\tablenotemark{a}}&
\colhead{Ref.}}
\startdata
SGR~1806$-$20&
          18 05 40&    $-$20 26&     $1\arcmin\times6\arcmin$& 1\nl
\tablevspace{1mm}
X-ray burst&
          18 05 40&    $-$20 25&     $2\arcmin\times4\arcmin$& 2\nl
AX~1805.7$-$2025&
          18 05 41&    $-$20 25 07&   1\arcmin& 2\nl
\tablevspace{1mm}
RX~J1808.6$-$2024&
          18 05 41.56& $-$20 25 03.4& 11\arcsec& 3\nl
\tablevspace{1mm}
VLA-HR &  18 05 41.68& $-$20 25 12.5& 0\farcs4& 4\nl
VLA-LR &\pha    41.76&\phd      13. & 1\farcs0& 5\nl
\tablevspace{1mm}
Star I\tablenotemark{b}&
          18 05 41.54& $-$20 25 11.6& 1\farcs0& 6,i\nl
Star A\tablenotemark{b}&
        \pha    41.68&\phd      12.1& 1\farcs2& 6,J,K\nl
Star 1 &\pha    42.02&\phd      11.0& 1\farcs0& 6,i\nl
Star 2 &\pha    42.01&\phd      07.3& 1\farcs0& 6,i\nl
Star 3 &\pha    42.29&\phd      13.8& 1\farcs0& 6,i\nl
Star 4 &\pha    41.97&\phd      21.5& 1\farcs0& 6,i\nl
Star 5 &\pha    41.97&\phd      18.2& 1\farcs0& 6,i\nl
Star 6\tablenotemark{b}&
        \pha    41.77&\phd      15.7& 1\farcs2& 6,J,K\nl
Star 7 &\pha    41.21&\phd      19.7& 1\farcs4& 6,K\nl
Star 8 &\pha    42.16&\phd      21.3& 1\farcs4& 6,K\nl
Star 9 &\pha    42.12&\phd      08.1& 1\farcs4& 6,K\nl
Star 10&\pha    41.45&\phd      09.5& 1\farcs4& 6,K\nl
Star 11&\pha    41.59&\phd      10.1& 1\farcs4& 6,K\nl
Star 12&\pha    41.73&\phd      21.7& 1\farcs4& 6,K\nl
\tablenotetext{a}{
   $r_{90}$ is the 90\% confidence error radius (or region).  That for
   VLA-HR is better than indicated but was increased to account for the
   complexity of the source. Likewise for VLA-LR. The relative
   uncertainties for stars determined from the same band is about
   0\farcs1 in each coordinate.}
\tablenotetext{b}{
   The position of Star~I as determined from the Guide Star Catalog
   astrometry has end figures 41.56 and 11.8.  Star~A has the same
   position in J and~K.  Star 6 has end figures 41.76 and 15.6 in~K.}
\tablerefs{           (1) Hurley et al.\ 1994;
                      (2) Murakami et al.\ 1994;
                      (3) Cooke 1993;
                      (4) Vasisht, Frail \&\ Kulkarni 1994;
                      (5) Kulkarni et al.\ 1993a;
                      (6) this work; i, J, K indicates the band in which
                          the position was determined.
}
\end{planotable}
\clearpage
\begin{figure*}
{\centering\leavevmode\epsfxsize\textwidth\epsfbox{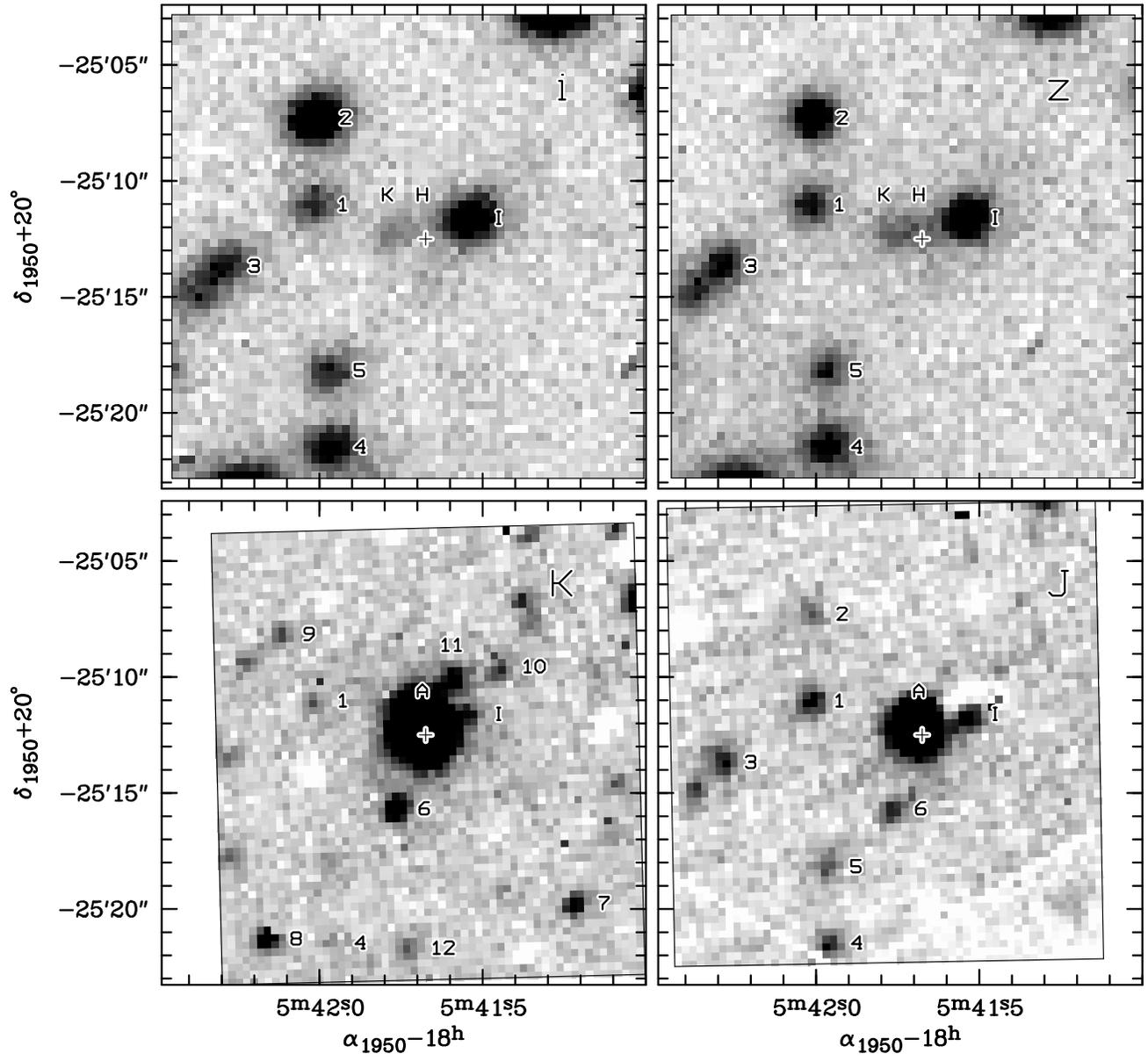}}
\caption{Optical (Gunn $i$ and $z$) and infrared (J and K band )
images of the field surrounding SGR~1806$-$20. The position of the
radio counterpart as determined from high-resolution VLA observations
is indicated with a cross.  Stars for which positions were determined
are indicated with a number or character to the right or over the
stellar image.}
\end{figure*}
\vfill\clearpage
\end{document}